\documentclass{aa}

\usepackage{graphicx}
\usepackage{natbib}
\usepackage{txfonts}
\usepackage{epstopdf}

\bibpunct{(}{)}{;}{a}{}{,}    

\def\apjs{ApJS}

\def\aap{Astron. Astrophys.}

\def\apj{Astrophys. J.}

\def\solphys{Sol. Phys.}
     
\def\nat{Nature}

\def\mnras{Mon. Not. R. Astron. Soc.}

\begin{document}
\title{NLTE modeling of Stokes vector center-to-limb variations in the CN violet system}
\author{A. I. Shapiro \inst{1}  \and D. M. Fluri \inst{2} \and S. V. Berdyugina \inst{3} \and M. Bianda \inst{2,4} \and R. Ramelli \inst{4}}
\offprints{A.I. Shapiro}

\institute{Physikalisch-Meteorologishes Observatorium Davos, World Radiation Center, 7260 Davos Dorf\\
\email{alexander.shapiro@pmodwrc.ch}
\and Institute of Astronomy, ETH Zurich, 8092 Zurich, Switzerland
\and Kiepenheuer-Institut fur Sonnenphysik, Schoneckstrasse 6, 79104 Freiburg, Germany
\and Istituto Ricerche Solari Locarno, Via Patocchi, 6605 Locarno-Monti, Switzerland} 

\date{Received 6 November 2008; accepted 23 February 2011}

\abstract 
{The solar surface magnetic field is connected with and even controls most of the solar activity phenomena. Zeeman effect diagnostics allow for measuring only a
small fraction of the fractal-like structured magnetic field. 
The remaining hidden magnetic fields can only be  accessed with the Hanle
effect.}
{Molecular lines are very convenient for applying the Hanle effect diagnostics thanks to the broad range of magnetic sensitivities in a narrow spectral region. With the UV version of the Zurich Imaging Polarimeter ZIMPOL II installed at
the 45 cm telescope of the Istituto Ricerche Solari Locarno (IRSOL), we simultaneously
observed intensity and linear polarization center-to-limb variations in two spectral regions containing the (0,0) and (1,1) bandheads of the CN $B \, {}^{2} \Sigma - X \, {}^{2}\Sigma$ system. Here we present an analysis of these observations.}
{We have implemented coherent scattering in molecular lines into a NLTE radiative transfer code. A two-step approach was used. First, we separately solved the statistical equilibrium equations and compute opacities and intensity while
neglecting polarization. Then we used these quantities as input for calculating scattering polarization and the Hanle effect.}
{We have found that it is impossible to fit the intensity and polarization simultaneously  at different limb angles in the framework of standard 1D modeling.  The atmosphere models that provide correct intensity center-to-limb variations fail to fit linear polarization center-to-limb variations due to lacking radiation field anisotropy.   We had to increase the anisotropy by means of a specially introduced free parameter.  This allows us to successfully interpret our observations. We discuss  possible reasons for underestimating the anisotropy  in the 1D modeling.}
{}


\titlerunning{NLTE modeling of the CN violet system}

\maketitle
%

\section{Introduction}\label{sec:intro}
The $Q/I$ spectrum formed by coherent scattering is a rich source of information about the solar atmosphere and was called therefore the ``second solar spectrum'' \citep{ivanov1991}. The properties of scattering processes are modified in a magnetic field via the Hanle effect \citep{stenflo1982}, and it makes the second solar spectrum a unique  tool to access weak, spatially unresolved entangled magnetic fields with mixed polarity, which cover about 99\% of the photospheric volume and cannot be measured with Zeeman effect because of signal cancellation \citep{stenflo2004, trujilloetal2004}.

Scattering in molecular lines plays an important role in forming the second solar  spectrum and even dominates in some spectral regions. The sensitivity of molecular lines to the Hanle effect  varies significantly with the total angular momentum of the upper level $J$. Since a narrow spectral
region can contain lines with quite different $J$ values, this allows us to employ the differential Hanle effect \citep[cf. ][]{stenfloetal1998,trujillo2003c,berdyuginafluri2004}. 

\citet{shapiroetal2007a}  developed the theory of the Hanle effect in the CN violet ($B \, {}^{2} \Sigma - X \, {}^{2}\Sigma$) system, which is a significant feature of the solar spectrum, including $Q/I$, and is very convenient for applying the differential Hanle effect diagnostics. Then, while employing a simple radiative transfer model, \citet{shapiroetal2007b} applied this theory for modeling CN violet system lines in the second solar spectrum and for determining the turbulent magnetic field strength. Despite the successful fit in several spectral regions, this simple model failed to fit strong CN lines. Moreover, it was impossible to self-consistently interpret the differences in the deduced model parameters obtained from different spectral regions. To solve these problems, we present here a new, more self-consistent model of 1D polarized
radiative transfer.

For our purpose, we carried out simultaneous observations of center-to-limb variations in the intensity and linear polarization  in the CN (0,0) and (1,1) bands. Such observations are very useful for constraining the atmosphere model and studying the height dependence of the deduced parameters as observations at different limb angles sample different heights in the solar atmosphere \citep{faurobertarnaud2003}. We developed a new model to interpret these observations. Here we present the first results of this interpretation and discuss the general problems of simultaneous modeling of $I/I_{\rm c}$ and $Q/I$ in 1D .

In Sect.~\ref{sec:obs} we briefly describe our new observations.  Then, in Sect.~\ref{sec:RT} we introduce the radiative transfer model and report our main assumptions.  In Sect.~\ref{sec:modeling} we discuss the problem of lacking radiation field anisotropy arising in simultaneous 1D modeling of $I/I_{\rm c}$ and $Q/I$  and show that it is not possible to fit all observed data in the framework of a 1D one- or multi-component atmosphere model. In Sect.~\ref{sec:lack} we discuss possible physical mechanisms leading to the insufficient anisotropy in 1D modeling and introduce a mean to increase the anisotropy  and solve the problem. Finally in Sect.~\ref{sec:newmodeling} we show that modeling with an increased anisotropy can provide us with a good fit quality and allows us to measure the magnetic field strength. We also discuss the model dependence of the deduced magnetic field value.

\section{Observations}\label{sec:obs}
With the UV version of the Zurich Imaging Polarimeter ZIMPOL II installed at the 45 cm telescope of the Istituto Ricerche Solari Locarno (IRSOL) we simultaneously observed Stokes $I$ and $Q/I$ center-to-limb variations in two wavelength regions, which contain the (0,0) and (1,1) bandheads.
Observations of these two regions were made on December 13 and December 14, 2007, respectively. For all observations the slit subtended 3.4 arcmin and was placed parallel to the nearest limb. The slit width corresponded to a spatial resolution of 1 arcsec.  Exposure time for one exposure frame was 5 sec, but to reach the required number of photons in order to get the desired signal to noise, the total exposure time at different $\mu$ positions was varying from 20 minutes to 48 minutes.
 Such a choice of the wavelength regions is justified for magnetic field diagnostics as bandhead regions contain a mixture of lines with different $J$-numbers and accordingly different magnetic sensitivity and formation height. Therefore, the differential Hanle effect technique can be applied. Moreover the bandhead is formed higher in the atmosphere than the radiation at the nearest wavelengths (1--2 {\AA} away from the bandhead) so the information about conditions at a broad range of heights can be inferred.

The intensity and linear polarization in the region containing the (1,1) bandhead (3868.9--3871.4 \AA) was observed at five limb angle positions from  $\mu=0.1$ to $\mu=0.5$ (here $\mu$ is the cosine of the angle between the propagation direction of radiation and the local solar radius). The intensity in the second region containing the (0,0) bandhead (3879.7--3883.6 \AA) was observed at ten limb angle positions from $\mu=0.1$ to $\mu=1.0$ with the constant step $\Delta \mu=0.1$. The polarization in the second region was observed at two positions:  $\mu=0.1$ and $\mu=0.2$.
The pointing of the telescope allows a precision of about 1 arcsec, but due to seeing and the curvature of the solar limb relative to the straight spectrograph slit, we need to take into account an uncertainty of about $\pm 0.02$ at $\mu = 0.1$. The error for $\mu$ values quickly decreases toward the disk centre.

{ The zero level of the polarization scale  is uncertain  in the observations. In Sect.~\ref{subsubsec:LSF} we present an algorithm which allows us to simultaneously determine the continuum polarization and the zero level of polarization scale. }

\section{Radiative transfer model}\label{sec:RT}
In the following we introduce our numerical method employed for solving the radiative transfer problem. We have performed calculations in a framework of plane-parallel model atmospheres. The modeling consists from two main steps, as was previously suggested by \citet{flurietal2003}. We firstly calculate opacities and intensity without taking into account any polarization. Then we iteratively calculate polarization assuming that opacities obtained in the first step remain unchanged.

\subsection{Phase matrix}
In the weak-field regime the redistribution matrix can be factorized into a scalar function depending only on frequencies of incoming and outgoing radiation and the phase matrix. The phase matrix does not directly depend on frequencies, but it has different appearance in different frequency domains \citep{bommier1997b, flurietal2003}.

Under the assumption of complete frequency redistribution (CRD) and an isotropic single-value turbulent magnetic field  the phase matrix for the line core domain (roughly speaking in this domain  both absorption and emission of a photon occur close to the line core) is given by \citep{flurietal2003}
\begin{equation}
{\vec P(\mu, \mu', W_2, W_{\rm H})} =
{\vec E}_{11}+3/4 \, W_2 W_{\rm H} \cdot {\vec P}^{(2)}, 
\label{eq:Pcore}
\end{equation}
and in the line wing domain by
\begin{equation}
{\vec P(\mu, \mu', W_2)} =
{\vec E}_{11}+3/4 \, W_2  \cdot {\vec P}^{(2)}, 
\label{eq:Pwings}
\end{equation}
In these equations $\mu$ and $\mu'$ indicate the outgoing and incoming directions of the scattered photon, $W_2$ is the effective scattering polarizability, $W_{\rm H}$ is the Hanle depolarization factor, and $\,{\vec E}_{11}$ and ${\vec P}^{(2)}$  are $2 \times 2$ matrices  (as we solve the radiative transfer problem for Stokes $I$ and $Q$ only) which are given by

\begin{equation}
{\vec E}_{11}=\left (
\begin{array}{cc}
1 & 0 \\
0 & 0 
\end{array}
\right ),
\label{eq:E11}
\end{equation}
and
\begin{equation}
{\vec P}^{(2)}(\mu, \mu')=\frac{1}{2} \left (
\begin{array}{cc}
\frac{1}{3} (1-3 \mu^2) (1-3 \mu'^2) & (1-3 \mu^2) (1-\mu'^2) \\
(1- \mu^2) (1-3 \mu'^2) & 3 (1- \mu^2) (1- \mu'^2) 
\end{array}
\right ).
\label{eq:P2}
\end{equation}

In agreement with  the principle of spectroscopic stability, line wings are not affected by the Hanle effect, and the Hanle depolarization factor $W_{\rm H}$ does not enter Eq.~(\ref{eq:Pwings}). Values of the dimensionless frequency for  a core-wing boundary separation  
are calculated for certain values of the Voigt parameter $a$ by \citet{bommier1997b}. For a given damping parameter $a$, we  interpolate
between the known values.

The matrix $\vec{E}_{11}$ represents isotropic, unpolarized scattering, while ${\vec P}^{(2)}$ describes linearly polarized coherent scattering \citep[e.g.,][]{stenflo1994}. Due to the latter even initially unpolarized radiation becomes polarized after scattering, and the degree of  polarization scales with the effective polarizability $W_2$.

The Hanle effect in the line core for the case of a turbulent single-value magnetic field is introduced by \citep{stenflo1982}
\begin{equation}
W_{\rm H}=1-0.4  \left (        \frac {   \gamma_{\rm H}^2   } {1+\gamma_{\rm H}^2} +   \frac { 4  \gamma_{\rm H}^2   } {1+4 \gamma_{\rm H}^2}            \right ),
\label{eq:WH}
\end{equation}
where
\begin{equation}
\gamma_{\rm H}=0.88 \, \frac{g_{\rm L} B}{\Gamma_{\rm R} + \Gamma_{\rm I} + d \cdot \Gamma_{\rm E}}.
\label{eq:gamma}
\end{equation}
Here $\Gamma_{\rm R}, \Gamma_{\rm I}$ and $\Gamma_{\rm E}$ are the radiative damping, inelastic and elastic collision rates, respectively, $d$ is the efficiency of the depolarization by elastic collisions,  $B$ is the magnetic field strength, and $g_{\rm L}$ is the effective  Land\'e factor of the transition's upper state.

{ 
\subsection{Source function}\label{subsec:SF}
The total source function $\vec{S} \equiv \vec{S} (\tau, \nu, \mu)$ consists from parts arising due to thermal and scattering processes in the continuum and spectral lines. It is given by
\begin{equation}
{\vec{S}}  ={ \left ( \alpha_{\rm c} {\vec{S}}_{\rm c}   +\sum\limits_{i=1}^{n} \alpha_{\ell}^i \varphi^{i} (\nu)  {\vec{S}}_{\ell}^i \right ) }/{ \left (  \alpha_{\rm c}+ \sum\limits_{i=1}^{n} \alpha_{\ell}^i \varphi^{i} (\nu)  \right ) },
\label{eq:totalS}
\end{equation}
where $\alpha_{\rm c}=k_{\rm c}+\sigma_{\rm c}$ is the continuum opacity (here $k_{\rm c}$ and $\sigma_{\rm c}$ are the continuum absorption and scattering coefficients), $\alpha_{\ell}^{i}=k_{\ell}^{i} + \sigma_{\ell}^{i}$ is the total line opacity in the {\it i}-th line (here $k_{\ell}^{i}$ and ${\sigma}_{\ell}^{i}$ are the line absorption and scattering coefficients), and $\varphi^{i} (\nu)$ is the normalized  Voigt profile function. 
$\vec{S}_{\rm c}   \!\! \equiv  \!\! \vec{S}_{\rm c} (\tau, \nu, \mu)$ is the continuum source function, and  $\vec{S}_{\ell}^{i} \! \equiv \! \vec{S}_{\ell}^{i} (\tau, \nu, \mu)$ is the line source function of the {\it i}-th line.

\subsubsection{Continuum source function}\label{subsubsec:CSF}
The continuum source function is given by
\begin{equation}
{\vec{S}}_{\rm c} = {\left (     k_{\rm c}   {   \vec B_{\rm th}      }  +    \sigma_{\rm c}     {\vec{S}}_{\rm c} ^{\rm sc}  \right )     }      / {  \left (    k_{\rm c} + \sigma_{\rm c}  \right )   }.
\label{eq:contS}
\end{equation}
Here the vector   $\vec{B}_{\rm th}=(B_{\nu},0)$ is the thermal source term, which is given by the Planck function. The monochromatic scattering in the continuum is introduced by
\begin{equation}
{\vec{S}}_{\rm c} ^{\rm sc} \equiv   {\vec{S}}_{\rm c}^{\rm sc} (\tau, \nu, \mu)  = \frac{1}{2} \int\limits_{-1}^1  {\vec P}  (\mu, \mu', W_2^{ }=W_2^{\rm eff}) {\vec I} (\tau, \nu, \mu') d \mu',
\label{eq:cont_scat}
\end{equation}
where ${\vec I}$ is the Stokes vector of the radiation field,  and  ${\vec P}$ is the phase matrix, which is given by Eq.~(\ref{eq:Pwings}).  The coefficient $W_2^{\rm eff}$ is the effective continuum polarizability.

According to Eqs.~(\ref{eq:Pwings}) and (\ref{eq:cont_scat})  the continuum polarization is proportional to the effective continuum polarizability. The exact value of the latter depends on many factors which are model dependent \citep[see][]{stenflo2005}. The situation is further complicated by the line blanketing effect. To modify the continuum  absorption coefficient $k_{\rm c}$ we used the value of the opacity correction factor from \citet{busaetal2001}, which was adjusted to better fit the intensities. 
Due to the lack of additional information we employed the same value to modify the continuum scattering coefficient $\sigma_{\rm c}$, which is not necessarily correct. The possible error will alter the scale of the continuum polarization. Therefore, the continuum effective scattering polarizability $W_2^{\rm eff}$  was considered as a free parameter noting however that it also contains a possible error from the line blanketing coefficient. 

The $W_2^{\rm eff}$ value is determined by fitting the observed center-to-limb variations of the continuum polarization. However to obtain the absolute value of polarization, the observed  $Q/I$ profiles have to be shifted by the value of the zero level of polarization scale   (see Sect.~\ref{sec:obs}).    We choose the shift for the observed $Q/I$ and the  $W_2^{\rm eff}$ value  so that the observed continuum polarization level coincides with the calculated. 
While the value of the shift is the same  for all $\mu$-values, the influence of the effective polarizability on the continuum polarization depends on $\mu$. As we have observations at several $\mu$-values, these parameters can be  determined. The obtained value of the effective continuum polarizability is about $W_2^{\rm eff}=0.6$ for both spectral regions.

\subsubsection{Line source function}\label{subsubsec:LSF}
For the photon to be emitted in the  {\it i}-th line, the upper level of this line has to be excited.  The excitation can be caused by the collisions or by the absorption of the photon in any {\it j}-th line line which shares the upper level with the  {\it i}-th line (see Fig.~\ref{fig:scheme}). Each excitation possibility gives rise to a separate term in the line source function. So the source function in the {\it i}-th line  within the CRD approximation can be written as
\begin{equation}
{\vec{S}}_{\ell}^{i}  = \sum\limits_{j} \varepsilon_{\rm sc}^{i,j}  \,  {\vec{S}_{i,j}^{\rm sc}} + \varepsilon_{\rm th}^{i} \vec{B}_{\rm th},
\label{eq:lineS}
\end{equation}
where $\varepsilon_{\rm sc}^{i,j} $ and $\varepsilon_{\rm th}^{i}$ are the branching coefficients (which are depth dependent) and ${\vec{S}_{i,j}^{\rm sc}}$ is the scattering integral which corresponds to the absorption in the {\it j}-th line and emission in the  
{\it i}-th line. It is given by
\begin{equation}
{\vec{S}_{i,j}^{\rm sc}} = \frac{1}{2}  \int\limits_{0}^{+\infty} \int\limits_{-1}^{+1}  
{\vec P} (\mu, \mu', W_2^{i,j}, W_H^{i}) {\vec I}(\nu',\mu') \varphi^j (\nu') 
\, d \mu'  d \nu'.
\label{eq:lineSsc}
\end{equation}
In  Eq.~(\ref{eq:lineS}) scattering integrals ${\vec{S}_{i,j}^{\rm sc}}$ with $i \neq j$  describe the Raman scattering, while the scattering integral  ${\vec{S}_{i,i}^{\rm sc}}$ describes the Rayleigh scattering. The last term corresponds to the thermal photons, emitted after the collisional excitation.

To illustrate the connection between the anisotropy of the radiation field and scattering polarization it is useful to express the scattering integrals in the form
\begin{equation}
 {\vec{S}_{i,j}^{\rm sc}}     =   \int\limits_{0}^{+\infty} \vec{\xi}^{i,j} (\nu')  \varphi^j (\nu')    \, d\nu',
\label{eq:lineSsep}
\end{equation}
}
where
\begin{equation}
\vec{\xi}^{i,j}(\nu') \equiv \left (
\begin{array}{c}
\xi_{\, \, I}^{i,j} (\nu') \\ 
\xi_{\, Q}^{i,j} (\nu')	
\end{array}
\right )=\frac{1}{2} \int\limits_{-1}^{+1}  
{\vec P} (\mu, \mu', W_2^{i,j}, W_H^{i}) {\vec I} (\nu', \mu') \, d \mu'.
\label{eq:xidef}
\end{equation}
Substituting the phase matrix from the Eq.~(\ref{eq:Pcore})--(\ref{eq:P2}) we obtain
\begin{equation}
\xi_{I}^{i,j} (\nu') = J_0^0(\nu') + \frac{\sqrt{2}}{4} W_2^{i,j} W_{\rm H} (3 \mu^2-1) \cdot J_0^2(\nu'), \\
\label{eq:xiI}
\end{equation}

\begin{equation}
\xi_{Q}^{i,j} (\nu')	= \frac{3\sqrt{2}}{4} \cdot W_2^{i,j} W_{\rm H} (\mu ^2-1)  \cdot J_0^2(\nu'),	
\label{eq:xiQ}
\end{equation}
where $J_0^0(\nu')$ is the mean intensity
\begin{equation}
J_0^0 (\nu') = \frac{1}{2} \int\limits_{-1}^{+1} I (\nu', \mu') \, d \mu',
\label{eq:jx}
\end{equation}
and $J_0^2(\nu')$ is the second moment 
\begin{equation}
J_0^2(\nu') \! \! = \!\! \frac{1}{4 \sqrt{2}} \! \int\limits_{-1}^{+1} \!\!
\left ( (3 \mu'^2-1) I (\nu', \mu') \! + \! 3 (\mu'^2-1) Q(\nu', \mu') \right )  d \mu'.
\label{eq:qx}
\end{equation}
The first term in Eq.~(\ref{eq:xiI}) describes isotropic scattering and depends only on the mean intensity. The second term represents coherent scattering and depends on the magnetic field. It's value is negligible in comparison with the first term, because  the anisotropy of the radiation field is usually very small. Therefore, Stokes $I$ is mainly determined by the mean intensity. In contrast, the polarization part of the line source function (see Eq.~(\ref{eq:xiQ})) does not contain contributions from  isotropic scattering and linearly depends on $J_0^2$. Due to the $W_{\rm H}$ factor this part is also subject to the Hanle effect.

Therefore, the $Q/I$-signal is proportional to the anisotropy of the radiation field which is given by 
\citep[see][]{trujillo2001, holzreuteretal2005}
\begin{equation}
A=\frac{J_0^2}{J_0^0}.
\end{equation}
By definition, the anisotropy describes the excess of the radiation incident from the radial directions (close to vertical) in comparison with the radiation incident from the side  (horizontal direction). 

\subsection{Two-step approach}
For solving the statistical equilibrium and the radiative transfer equations  for the CN violet system lines we employ the numerical NLTE code (hereafter RH-code) written by \citet{uitenbroek2001}  and based on the multilevel accelerated lambda iterations (MALI) method \citep[cf.][]{rybickihummer1991, rybickihummer1992}. Contributions from other lines were considered as ``background'' and treated in LTE.  The statistical equilibrium equations in the RH-code are solved under the assumption of LTE populations within a single vibrational level since it would be time-consuming to solve them for each individual vibrational-rotational level (as we have to deal with more than 2000 transitions simultaneously). 
Such an approximation  is commonly used in NLTE molecular calculations \citep[cf.][]{thompson1973, mountlinsky1974, ayreswiedemann1989, uitenbroek2000}, because oscillator strengths of pure rotational radiative transitions are negligibly small and radiative processes can contribute to the rotational population balance only via two-step processes (Raman scattering for the ground electronic state and emission followed by absorption for the exited state). 

We have included calculations of electronic-vibrational molecular transitions with the fine structure into the RH-code in order to use it for the CN violet system computation. Although the modified code can be used for a more general case, for the CN violet system we consider only diagonal vibrational bands as the Franck-Condon factors of the non-diagonal bands are much lower than those of the diagonal.

With the RH-code we compute the opacities and intensity, neglecting polarization. Then, we use these as the input for the second code (hereafter POLY-code) written by \citet{fluristenflo2003} and \citet{flurietal2003}, which iteratively solves the polarized radiative transfer equation, taking into account the Hanle effect and assuming that opacities obtained in the RH-code remain unchanged (which is a good approximation since the degree of polarization in our calculations is always lower than 1\%). The POLY-code was adjusted for dealing with many blended lines as initially it was designed for treating only a few non-overlaping atomic lines, while in molecular bands even a narrow spectral region can contain several hundred lines, which have to be computed simultaneously.

\subsection{Thermal-scattering branching of the source function}\label{sec:branching}
\begin{figure}
\resizebox{\hsize}{!}{\includegraphics{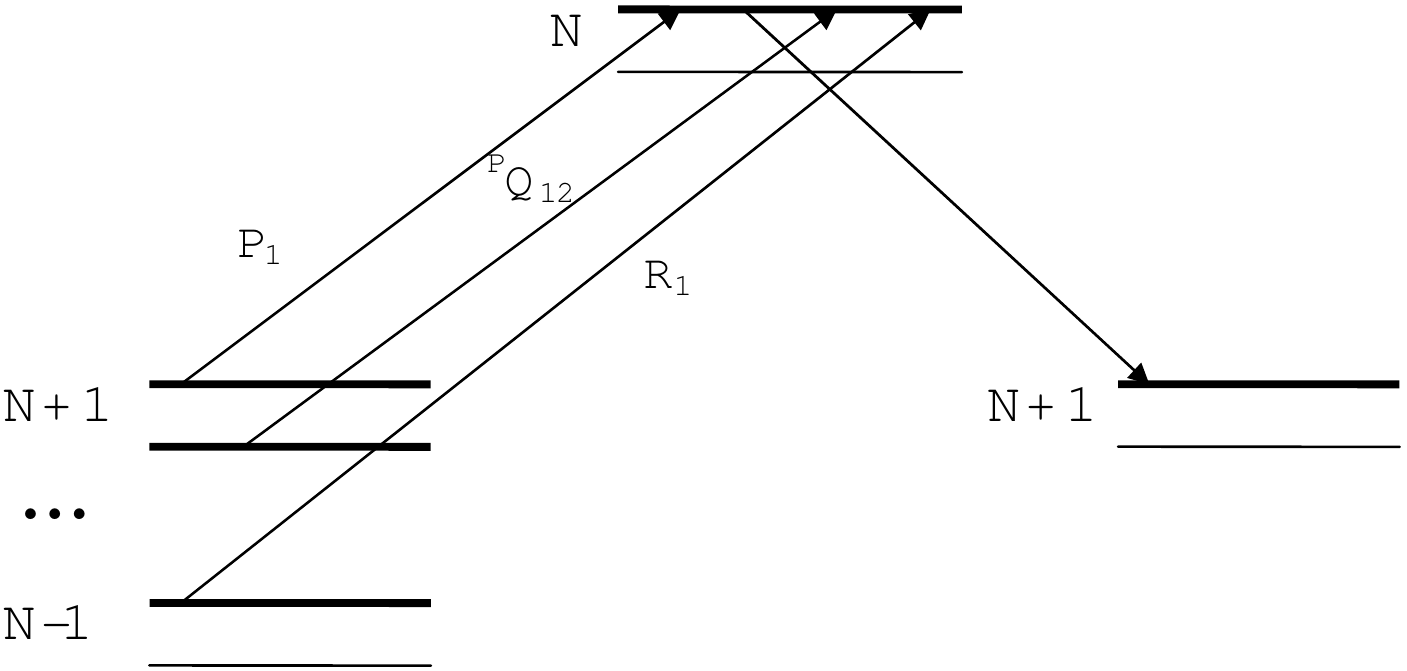}}
\caption{Scheme of the radiative excitation between the fine structure levels of the CN violet system. Doublet states are marked with the quantum number $N$. Allowed transitions in the $P_1$, $R_1$, and $^{P} Q_{12}$ branches are indicated.}
\label{fig:scheme}
\end{figure}

\begin{figure*}
\centering
\includegraphics{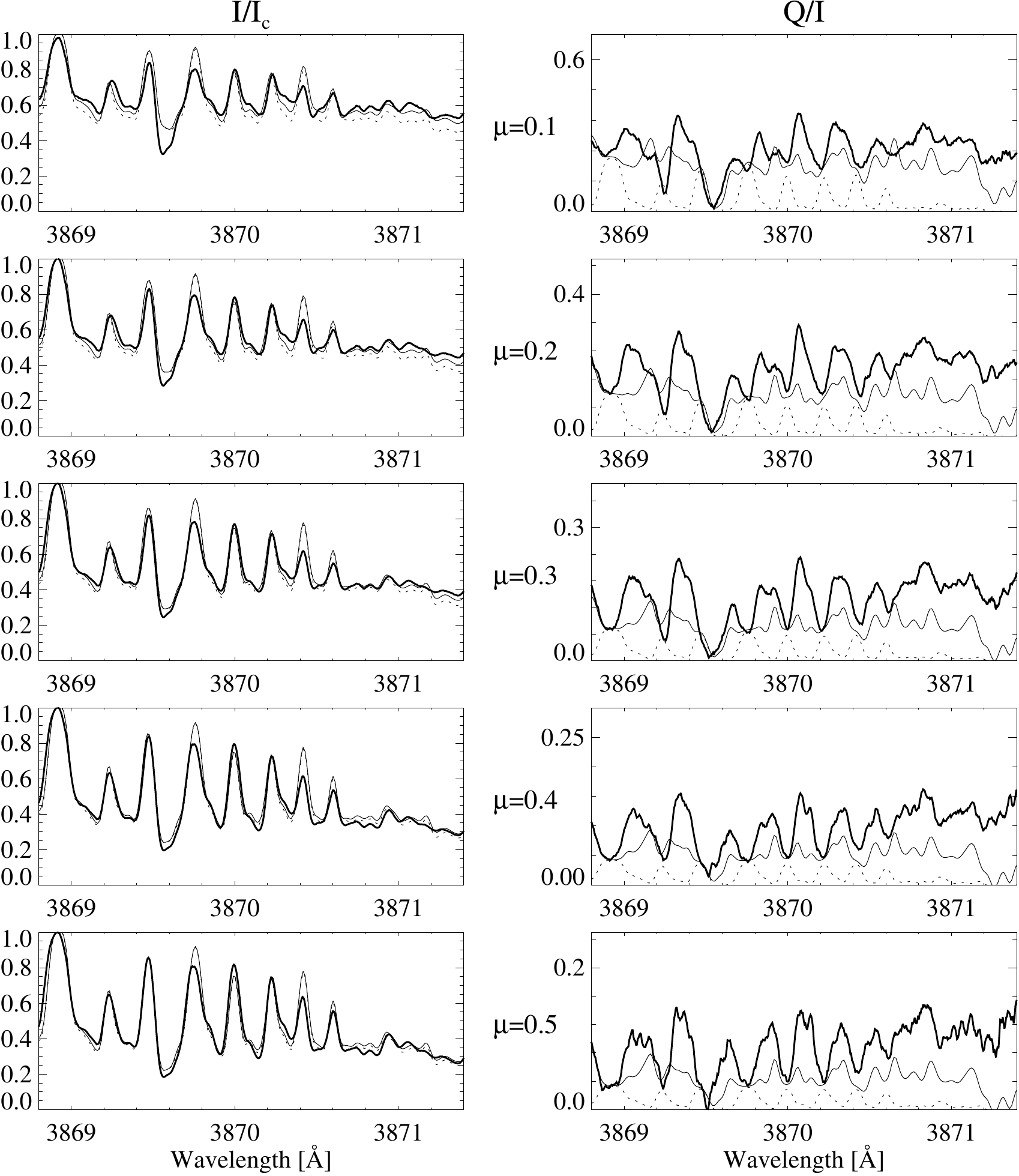}
\caption{Observations (thick solid lines) of Stokes $I/I_{\rm c}$ and $Q/I$  in the region of the (1,1) bandhead and fits with the FALC atmosphere model. The assumed magnetic field strength in the calculated spectra is 0 G. Two curves presented in each picture correspond to the NLTE (thin solid) and LTE (dotted) calculations.}
\label{fig:IQ3872}
\end{figure*}
As in most  NLTE-codes, the line source function in the RH-code is calculated as a function of  NLTE populations, temperature and frequency \citep[see][]{uitenbroek2000}. The source function deviations from the LTE are defined via  population departure coefficients (the ratio between the LTE and NLTE populations). Such a formalism is sufficient for calculations of the intensity field but does not provide any direct information about the contribution of scattering processes to the line.  However, for polarization calculations it is important to know the exact balance between scattering and thermal processes, since polarization is produced only by scattering. Therefore in the POLY-code the source function is calculated as a sum of scattering and thermal parts (see Eq.~(\ref{eq:lineS})). 

{  The conversion from the RH to the POLY-approach is carried out by choosing  branching coefficients $\varepsilon_{\rm sc}^{i,j}$ and $\varepsilon_{\rm th}^i$ in Eq.~(\ref{eq:lineS})}. Defining these coefficients is a standard problem for a two-level system, which is described in many textbooks \citep[e.g.][]{mihalas1970}. However, in the case of the CN violet system the situation becomes much more complicated as we have to consider a lot of levels, many of which are collisionally and radiatively coupled with each other. 

To find the branching coefficients  let us consider the prehistory of the photon which was emitted in the {\it i}-th line, when molecule changed from the upper to the lower  state. If the excited state  was collisionaly populated (it could be both the collisional excitation from a lower level or de-excitation from a higher energy level) a thermal photon,  which contributes {  to the last term of Eq.~(\ref{eq:lineS}), will be emitted. However, if it was radiatively populated by the absorption of the photon in the {\it j}-th line (which shares the upper level with  the {\it i}-th line), a nonthermal photon, which contributes   to the $\varepsilon_{\rm sc}^{i,j}  \,  {\vec{S}_{i,j}^{\rm sc}}$ term in  Eq.~(\ref{eq:lineS}), will be emitted. Therefore, if $C_{\rm tot}$ and $R_{\rm tot} \equiv \sum_j R_j$ are the total rates of the collisional and radiative processes populating the upper level, then the part of thermal photons in the {\it i}-th line is given by
\begin{equation}
\delta_{\rm th} \equiv C_{\rm tot}/(C_{\rm tot}+R_{\rm tot}),
\label{eq:th}
\end{equation}
while the other part will be nonthermal:
\begin{equation}
\delta_{\rm sc} \equiv R_{\rm tot}/(C_{\rm tot}+R_{\rm tot}).
\label{eq:sc}
\end{equation}
}

As we solve the statistical equilibrium equations only for vibrational levels, the coefficients $\delta_{\rm th}$ and $\delta_{\rm sc}$ are also calculated  only for  vibrational levels of the excited electronic state $B {}^2 \Sigma$  (adding up all significant radiative and collisional processes, populating the considered vibrational level). We assume that the same fractions of thermal and scattered photons  $\delta_{\rm th}({\rm v})$ and $\delta_{\rm sc}({\rm v})$ apply to every rotational level within a given rotational band (v,v). This seems to be a good approximation as the rotational energy is low in comparison with the electronic energy of the excited  $B {}^2 \Sigma$ state.

In the case of a two-level system the branching coefficients  $\varepsilon_{\rm th}^i$ {  and $\varepsilon_{\rm sc}^{i,i}$}  are equal to the  coefficients $\delta_{\rm th} $ and $\delta_{\rm sc} $. However in our case of a multi-level system this equality is not valid and the branching coefficients depend not only on the coefficients $\delta_{\rm th} $ and $\delta_{\rm sc} $, but rather on the whole balance of populating and depopulating rates.

The following algorithm was employed to compute {  the $\varepsilon_{\rm sc}^{i,j}$ and $\varepsilon_{\rm th}^i$ coefficients}. Using the NLTE populations of the $l_i$ and $u_i$ levels from RH, we firstly calculate the line source function with the standard equation
\begin{equation}
S_{\ell}^i=\frac{n_{u_i} A_{u_i l_i}}{n_{l_i} B_{l_i u_i} - n_{u_i} B_{u_i l_i}},
\label{eq:lineSst}
\end{equation}
where $A_{kk'}$ and $B_{kk'}$ are the Einstein coefficients.  The source function given by  Eq.~(\ref{eq:lineSst})  has to be  equal to the intensity part of the source function given by  Eq.~(\ref{eq:lineS}). {  So one can write
\begin{equation}
S_{\ell}^i= \frac{n_{u_i} A_{u_i l_i}}{n_{l_i} B_{l_i u_i} - n_{u_i} B_{u_i l_i}} = \varepsilon_{\rm th}^i {{B}}_{\rm th}  + \sum\limits_j  \varepsilon_{\rm sc}^{i,j}  I^{\rm sc}_{i,j},
\label{eq:th1}
\end{equation}
where $I^{\rm sc}_{i,j}$ is  the intensity part of the scattering integral from  Eq.~(\ref{eq:lineSsc}). 

Secondly,  the ratio between the thermal (arising from the collisional excitation) and {\it j}-th nonthermal (arising from the radiative absorption in {\it j}-th line) contributions to the source function is given by
\begin{equation}
\frac{\varepsilon_{\rm th}^i {{B}}_{\rm th}}{\varepsilon_{\rm sc}^{i,j}  I^{\rm sc}_{i.j}}=\frac{C_{\rm tot}}{R_{j}}.
\label{eq:th2}
\end{equation}
The Eq.~(\ref{eq:th2}) can be written for each from the nonthermal contributions to the source function and therefore, together with the Eq.~(\ref{eq:th1})  the number of the equations for the branching coefficients is equal to the number of the branching coefficients.   Since the equations are not linearly dependent the branching coefficients can be  uniquely determined.  We note that in our case of the  radiative transfer problem with the multi-level coupling the sum of the branching coefficients is not necessarily one (see the lower panel of Fig. 3).

For every line in the CN violet system there are three different ways to radiatively populate the upper state   before the emission (see Fig.~\ref{fig:scheme}). One of them, in a main branch, corresponds to  Rayleigh scattering and the two others to Raman scattering (one in a main  branch and one in a satellite branch). The oscillator strength of the transitions in a satellite branch are approximately four orders of magnitude weaker than the transitions in the main branches and therefore the satellite transitions can be neglected. So in our calculations the line source function consists from two scatterings terms (one corresponding  to the Rayleigh and another to the Raman scattering)  and one thermal term (see Eq.~(\ref{eq:lineS}) and the second equality in Eq.~(\ref{eq:th1})).}

\subsection{Collisional rates}
There are three main types of inelastic collisions with molecules. The first one alters only the total angular momentum number $J$ of the molecular state, the second also the vibrational state, and the third the electronic state. There are many theoretical approximations and parametrizations for the rates of these three collision types \citep[cf.][]{thompson1973,hinkle1975,ayreswiedemann1989}. The accuracy of these approximations is very low, and the parameters for the CN molecule are unknown. Moreover it is not possible to distinguish the elastic and inelastic collisions from our observations. Therefore, we neglect elastic collisions. It does not influence our conclusions but can slightly affect the value of the deduced magnetic field strength.

In our calculations we used the Landau-Teller formula for the relaxation time of the excited state \citep{ayreswiedemann1989}
\begin{equation}\label{eq:LT}
\ln{ (P_{\rm X} \, t_{\rm CN-X})} =A_{\rm X} T^{-1/3}-B_{\rm X},
\end{equation}
where $P_{\rm X}$ is the partial pressure of the collision partner ``X'', $t_{\rm CN-X}$ is the relaxation time of the excited CN state, $T$ is the temperature, and $A_{\rm X}$ and $B_{\rm X}$ are  free parameters. The associated collisional de-excitation rate per one molecule is
\begin{equation}\label{eq:C}
{\cal C}={  (t_{\rm CN-X} \, (1-\exp{(-\beta)})       }^{-1},
\end{equation}
where $\beta = \Delta E / kT $ is the excitation parameter.

As there are several collisional agents (most importantly neutral hydrogen and electrons) the collisional rates depend on many free parameters. However for computing the branching coefficient and scattering polarization only the total collisional rate (independently on collisional agent and type) is relevant (see Sect.~\ref{sec:branching}). We assume then that the total collisional rate can be also calculated with Eqs.~(\ref{eq:LT})--(\ref{eq:C}), using the partial pressure of neutral hydrogen (as the collisions with neutral hydrogen seems to be the strongest and moreover the depth dependence of the electron particle density is close to the one of neutral hydrogen in the region of the CN lines formation).

With such an approximation the collisional rate depends only on two free parameters: $A$ and $B$, which define the temperature and density dependence, respectively. At the CN line formation height the density drops much faster than the temperature. Therefore, the parameter $A$ only slightly affects the dependency of the collisional rates on height in the atmosphere, in particular also when considering the small exponent of the temperature in Eq.~(\ref{eq:LT}). 
We found that the best fit quality can be reached if we put the value of $A$ parameter equal to zero. Let us notice, however, that even under this approximation the branching coefficients $\delta_{\rm th}$ and $\delta_{\rm th}$ still depend on temperature due to the Einstein coefficients, which connect excitation and de-excitation collisional rates and affect the level populations.

Finally we end up with only one free parameter $B$ which defines the dependency of the collisional rate on height. For convenience we have replaced it with the $\delta_{\rm th}$ coefficient at the temperature minimum layer (hereafter $\delta_{\rm th}^{\rm min}$). Considering the strong limitations of the current collision theory with molecules, the unknown collision parameters for CN, and our need to know only the total collision rate such approximation becomes very practicle and improves the stability of the procedure to fit observations. We can control the collision rate with a single free parameter and nonetheless account for the height dependence via the Landau-Teller expression. In fact, by fitting observations we can even gain empirical constrains for collisional rates with the CN molecule.

\section{Center-to-limb variations with standard modeling}\label{sec:modeling}

\subsection{One-component atmosphere modeling}\label{subsec:one}
As a first step we performed computations in a frame of a plane-parallel one-component model atmosphere, namely model FALC by \citet{fontenlaetal1993}.
In Fig.~\ref{fig:IQ3872} both the LTE and NLTE calculations of the intensity and linear polarization for different limb angles in the region of the (1,1) bandhead are presented. 
The region contains about 100 CN lines and several atomic blends, among them two strong iron lines around 3869.5 \AA. These blends in principle 
have to be treated in NLTE too, but for simplicity we  treated all blends in LTE, which reduces the quality of the fit at wavelengths affected by blends but does not influence our conclusions.

From Fig.~\ref{fig:IQ3872} one can see that the calculated and observed linear polarization curves have completely different shapes.
Surprisingly, the calculated polarization {  dramatically decreases and for high $\mu$-values even  becomes  negative (perpendicular to the solar limb) in the bandhead at 3871.2 {\AA}, completely contradicting the observations.} Here we have chosen the collisional coefficient for the NLTE calculations $\delta_{\rm th}^{\rm min}$ equal to 0.15, so that the overall level of polarization outside the bandhead fits the observations. The overall  polarization level can be increased by choosing a smaller collisional coefficient, but this will make the situation at the bandhead polarization even worse as its absolute value will  increase but it will remain negative. Even excluding the particular bandhead region, the calculated shape of the $Q/I$ curve does not correspond to the observed one and it can not be corrected with the Hanle effect (see Sect.~\ref{sec:newmodeling}).

While the calculated polarization shows such a strange behavior, the quality of the intensity fit is good for all five $\mu$-values. In the LTE calculations lines are too deep, while taking into account the NLTE effects makes lines a bit weaker and leads to the good agreement with observations. It is interesting that the higher $\mu$-value, the weaker NLTE corrections. This is connected with the fact that observations at greater limb angles sample deeper layers of the solar atmosphere where the density and collision rates are higher and, therefore, the deviations from LTE become smaller.

Thus we find that the standard FALC model is able to reproduce the center-to-limb variations for the intensity but dramatically fails for polarization. The good fit to Stokes $I$ was expected because the FALC model has been empirically optimized to reproduce the solar intensity spectrum.

\begin{figure}
\resizebox{\hsize}{!}{\includegraphics{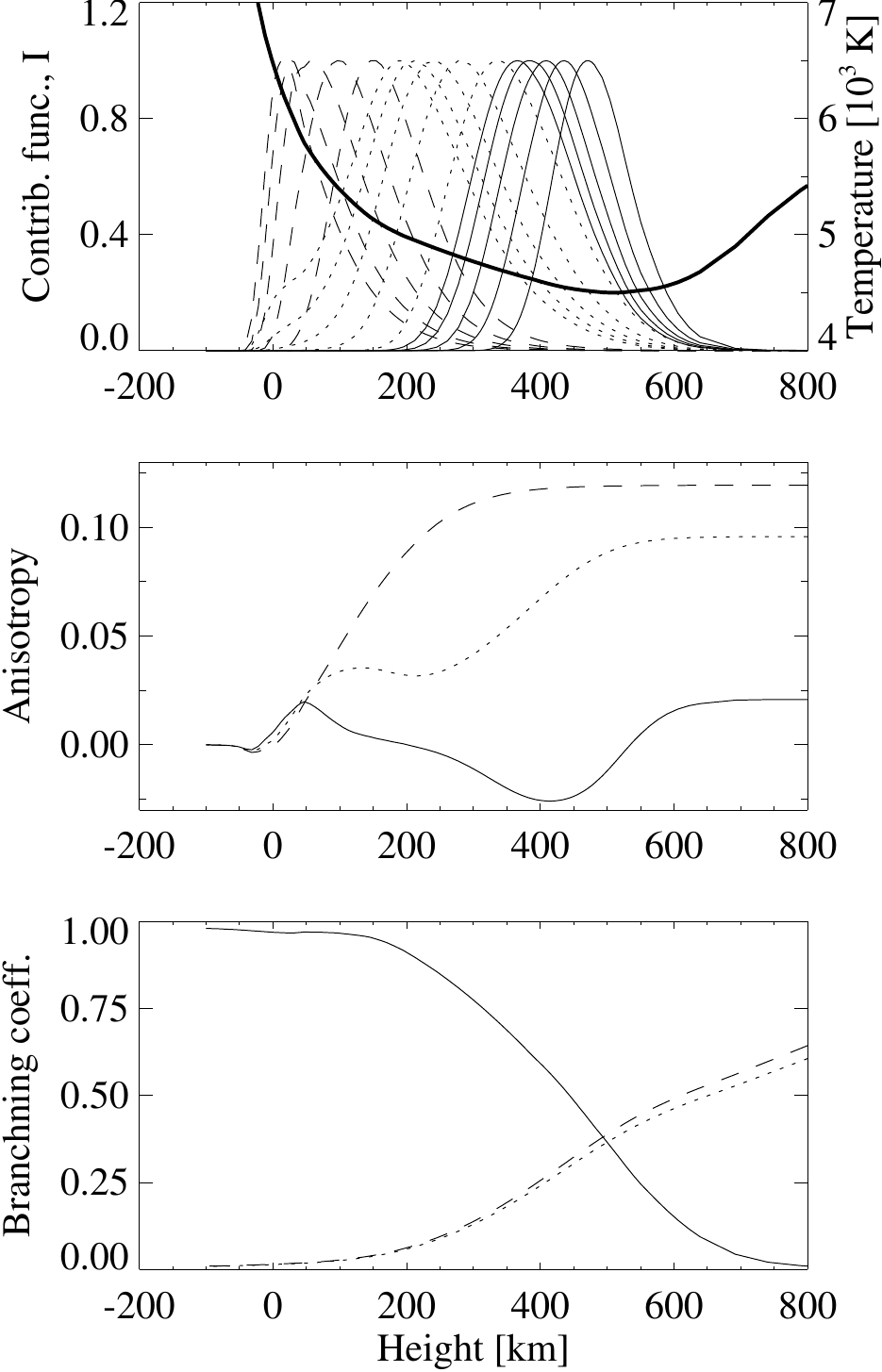}}
\caption{Upper and middle panels: Normalized contribution functions of Stokes $I$ as well as the anisotropy at 3871.2 {\AA} in the (1,1) bandhead (thin solid lines), at  3869.1 {\AA} in a CN blend (dotted lines) and at the local continuum (dashed lines). The contribution functions are given for  five $\mu$ values: from 0.1 (rightmost) to 0.5 (leftmost). The FALC temperature as a function of height is given in the upper panel (thick solid line, right scale). {  Lower panel: Branching coefficients for the source function of $P_1(14)$ (3869.1 \AA) line: thermal (solid line), Raman (dotted line) and Rayleigh (dashed line). Assumed  $\delta_{\rm th}^{\rm min}$ value is 0.3.} }
\label{fig:contr3872}
\end{figure}

In the following paragraphs we discuss the reasons why calculations with the FALC model atmosphere yield good results for the intensity spectrum but not for the second solar spectrum. The main point to recall is the fact that intensity and scattering polarization are influenced by different physical quantities. As discussed in Sect.~\ref{subsec:SF} scattering polarization scales with the anisotropy of the radiation field, which has negligible impact on Stokes $I$ in the case of the solar atmosphere. The discrepancy in the fits to the intensity and polarization spectra can thus be understood in the following terms: the FALC model correctly reproduces the mean intensity and thus Stokes $I$ within the solar atmosphere, but yields improper radiation field anisotropy and thus very bad fits to scattering polarization.

For a deeper understanding we briefly review the main factors that influence the anisotropy of the radiation field within a stellar atmosphere. The anisotropy mainly results from two competing contributions, namely the source function gradient and a surface effect \citep{trujillo2001, holzreuteretal2005}. The source function gradient is, under LTE conditions, due to the temperature gradient within the atmosphere. The temperature drop in the photosphere leads to limb-darkening. This means that at any point within the photosphere the upward moving radiation field is strongest in the vertical direction and reduces to more inclined directions from the side, which by definition contribute to positive anisotropy. The surface effect, on the other hand, is active only within one mean free photon path from the top of the atmosphere and affects the downward moving part of the radiation field. Within these topmost layers the optical path length between the top and some point within the atmosphere becomes significantly greater for inclined directions. As a result the downward directed radiation field is limb-brightened and thus contributes to negative anisotropy simply due to the presence of the surface.

The two effects lead to an anisotropy that in general depends on the height within the atmosphere and on wavelength because the source function gradient is height and wavelength dependent. Furthermore, the influence of the surface effect is zero deep in the atmosphere, reaches a maximum around an optical depth  $\tau=1$ and reduces towards the top because the absolute amount of downward radiation becomes negligible very close to the top.

Figure~\ref{fig:contr3872} illustrates both competing factors, their influence on anisotropy, and their subsequent effect on scattering polarization for the case of spectral modeling of the (1,1) band in the CN violet system based on the FALC atmosphere. From the contribution functions of Stokes $I$ we find that our observations sample a wide range of the solar atmosphere from the very bottom of the photosphere up to about 600 km. This is mainly due to the increased opacity when approaching the bandhead, which contains a large number of blended CN lines and thus forms at higher atmospheric layers. We consider anisotropy and contributions functions for Stokes $I$ and $Q$ for three representative cases, namely at 3871.2 {\AA} (bandhead), at 3869.1 {\AA} (CN blend), and for the local continuum.

The height dependence of the anisotropy at the bandhead wavelength clearly displays the role of the two competing effects. Deep in the atmosphere (below 0 km) the photons thermalize and anisotropy vanishes even in presence of a large-scale temperature gradient. Around a height of 400 km the surface effect dominates and causes a strongly negative anisotropy. At even higher layers, the surface effect diminishes and, as a consequence, the source function gradient effect manages to increases the anisotropy to low but positive values. The anisotropy remains low at the top because the temperature gradient and thus also the source function gradient are relatively small near the temperature minimum. In 1D modeling and for a given wavelength the anisotropy only varies with height. However, when observing closer to disk center we sample deeper parts of the atmosphere so that we get greater contributions from the layers with negative anisotropy. {  Therefore, with increasing $\mu$ value, the value of the emergent polarization becomes smaller and even negative  for high $\mu$ (see Fig.~\ref{fig:IQ3872}).}

The behavior of the anisotropy for the CN blend at 3869.1 {\AA} is in principle  similar to that in the bandhead, although the emitted radiation at this wavelength forms lower in the atmosphere, where the  
temperature gradient $dT/dh$ is larger. Also, the gradient in the optical depth scale $dT/d \tau$ is even higher, because of the smaller opacity at this wavelength. It is this gradient that defines the limb-darkening law and, therefore, important for the anisotropy. Hence, the anisotropy at 3869.9 {\AA} is always greater than that at the bandhead  and never becomes negative.

The continuum radiation forms much lower in the atmosphere, where the temperature gradient is very high so that the surface effect becomes negligible. Therefore, the behavior of the continuum anisotropy is simpler than in the previous cases, and it monotonically increases with height.

We considered a number of additional known models such as FALA, FALC, FALF and FALP of \citet{fontenlaetal1993}, FALX of \citet{avrett1995} and colder models like model 2 of \citet{anderson1989} and  AYCOOL of \citet{ayresetal1986} and \citet{solankietal1994}. Only the model AYCOOL which has a strong temperature drop in the highest layers is able to produce relatively well the overall level of polarization, but it gives an incorrect polarization ratio in the bandhead compared to lines 0.5--1 {\AA} away from it. This model gives also completely wrong intensity profiles.

The main conclusion from the above discussion is that the anisotropy of the radiation field driven by the  temperature gradient in the FALC model is too low to explain the observed polarization. On the other hand, increasing the temperature gradient, for example, by lowering the temperature in the chromosphere, 
fails to reproduce the observed center-to-limb variations in the intensity. Therefore, it appears that one-component modeling  can reproduce either $I/I_{\rm c}$ or $Q/I$ signals depending on the choice of the temperature gradient, but not both simultaneously.

{  As was discussed in Sect.~\ref{sec:branching}  the  source function of the {\it i}-th line contains two scattering terms: the Rayleigh one, which corresponds to the radiative excitation by the photon absorption in the same {\it i}-th line and the Raman one, which corresponds to the radiative excitation by the photon absorption in another {\it j}-th line (which shares the upper state with the {\it i}-th line, see Fig.~\ref{fig:scheme}). Therefore the Stokes $Q$  source function of the {\it i}-th line depends not only on the anisotropy of the radiation at the wavelength of the {\it i}-th line, but also on the anisotropy of the radiation at the wavelength of the {\it j}-th line. In some cases these lines can be up to about 50 {\AA} away from each other. It means that due to the Raman scattering the polarization signals at two relatively distant regions of the spectrum can be tightly connected with each other. This significantly complicates the calculations as for computing the polarization even in the narrow spectral region one has to calculate the intensity and the anisotropy basically for the entire CN violet system. This system is blended by the large amount of the strong atomic lines, many of those have to be calculated in NLTE \citep{shapiroetal2010}. For simplicity we calculated all the atomic lines in LTE but slightly adjusted their oscillator strengths.

The effect from the Raman scattering is especially strong in the case of the optically thick lines which presence significantly modifies the anisotropy of the radiation field. For example without the Raman scattering term the polarization at the bandhead (where the optical depth is the highest) would be even lower than shown in Fig.~\ref{fig:IQ3872} and negative for all $\mu$-values. However this strong negative polarization is diluted with the polarization brought from  the different regions of the spectrum by the Raman scattering.}

\subsection{Two-component atmosphere modeling}\label{subsec:multi}
Next we try to fit the observations with a multi-component atmosphere model. There have been many theoretical and observational evidences for a temperature bifurcation in the upper photosphere and lower chromosphere. One comes from observations of  CO fundamental vibration-rotation transitions in the near infrared  \citep[][and references therein]{noyeshall1972, uitenbroek2000}. Also, to  explain the difference between field   strengths deduced from Hanle effect analysis of the \ion{Sr}{i} 4607 {\AA}  and  ${\rm C}_2$ lines, \citet{trujilloetal2004} suggested that the observed polarization in these lines is formed in regions with different temperature and anisotropy \citep[see also][and references therein]{asensiotrujillo2005, trujilloshchukina2007}. Further, \citet{holzreuteretal2006} showed that a temperature bifurcation is necessary for the explanation of the linearly polarized \ion{Ca}{ii} K profile.

{  We tried to fit the polarization signal with the combination of the models listed in the Sect.~\ref{subsec:one}. However the differences between the anisotropies in the considered models are very small (except  AYCOOL model). The anisotropy have also a different sign in different parts of the solar atmosphere. We found that the combination of these atmosphere models also cannot reproduce the observed $Q/I$ curve. }

We tried also to construct new two-component atmosphere models. We found that it is possible to construct a model which can fit reasonably well  all five $Q/I$ curves but at the same time fails to fit $I/I_{\rm c}$ curves. We found another model which can fit simultaneously $I/I_{\rm c}$ and $Q/I$ curves for one particular $\mu$-value, but fails to fit observations for other $\mu$-values. So, we can conclude that although some compromises can be found, it is impossible to fit simultaneously the whole set of the observed data by combining two models.

\begin{figure*}
\centering
\includegraphics{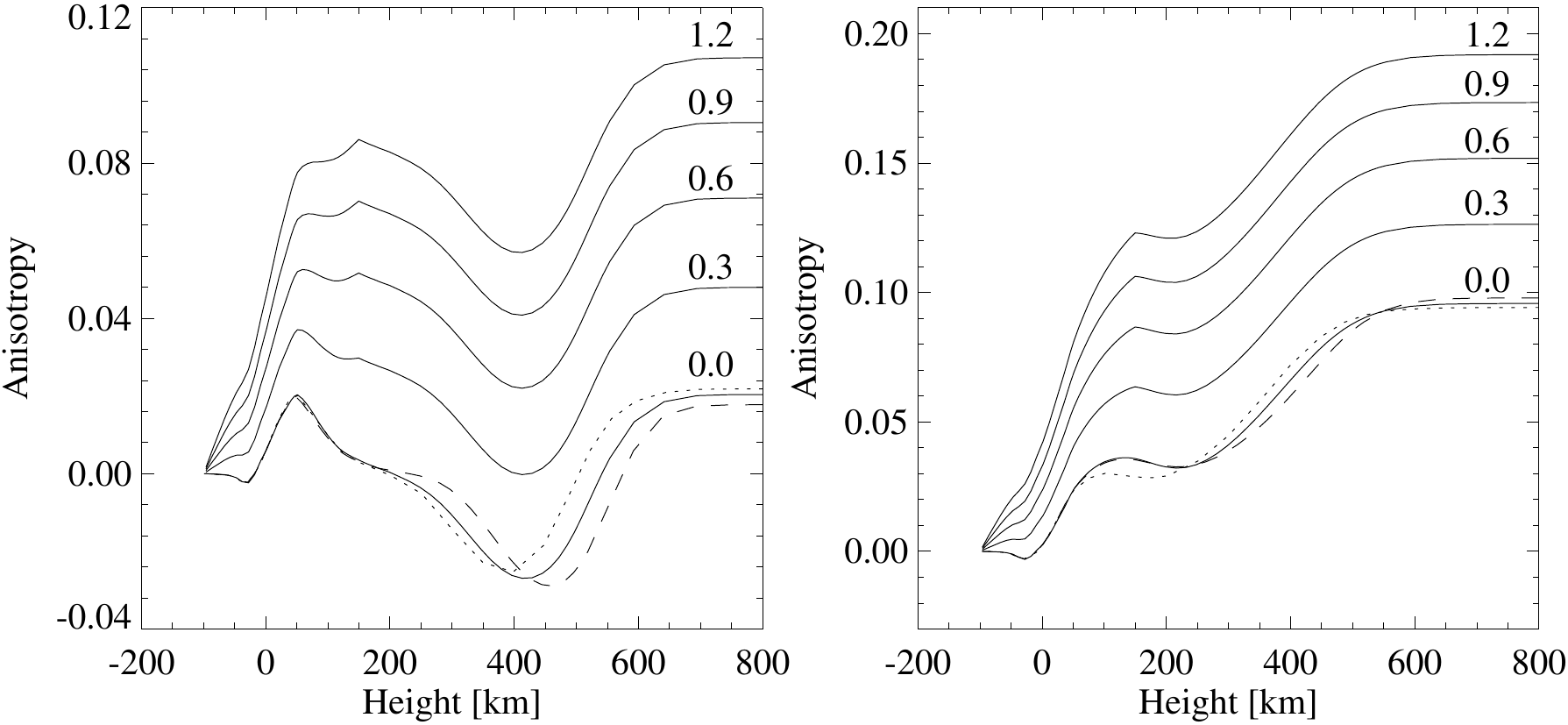}
\caption{The dependence of the anisotropy on the height calculated with the FALC model (solid lines) for several $f$ parameter values (marked near the curves), and  with the FALA (dotted line) and FALF (dashed line) models for $f=0$ (this implies  calculations, without anisotropy alteration, see Sect.~\ref{subsec:correction}). The anisotropy was computed at 3871.2 {\AA} in the (1,1) bandhead (left panel) and at 3869.1 {\AA} in a CN blend (right panel).}
\label{fig:anis}
\end{figure*}

The failure to fit simultaneously $I/I_{\rm c}$ and $Q/I$ signals  with the considered combinations of atmosphere models for the broad range of $\mu$-values is not surprising. To reproduce the polarization signal the temperature gradient in one of the atmosphere components has to be larger than in FALC (see Sect.~\ref{subsec:one}). However, such a large gradient will result in wrong intensity center-to-limb variations, so the temperature gradient in the second atmosphere component has to be lower than in FALC to compensate it, which in turn cause a very small or even negative polarization signal.  So the model which has to correct  center-to-limb variations in the intensity spoils the polarization center-to-limb variations. Introducing  atmosphere models with more than two components will not change the situation.

\section{Effect of horizontal temperature variations}\label{sec:lack}
\subsection{Additional anisotropy due to 3D effects}\label{subsec:1Dproblems}
The main problem in the interpretation of our observations is a lack of radiation field anisotropy. 
There are many reasons for  underestimating the anisotropy in the 1D modeling. Basically, each heterogeneity in the solar atmosphere will increase the anisotropy. Using  snapshots  from a hydrodynamical simulation of the solar surface convection  \citet{shchukinatrujillo2003} found indeed very strong horizontal fluctuations of the anisotropy  in the \ion{Sr}{i} line. It varies from  negative values up to about 15\% at the height of 400 km. 

One mechanism for increasing the anisotropy can be due to  horizontal temperature variations. Let us consider, for example, a two-component atmosphere model. In general it is incorrect to separately calculate Stokes $Q$ in each of the components and then linearly sum these individual contributions up to find the total emitted Stokes $Q$. 
The adjacent areas with different temperatures affect each other and change the anisotropy behavior. Indeed, the areas of the atmosphere which are described by the warm component are surrounded by colder areas of the second cold component. Therefore, close to the boundary the radiation coming from aside originates in the cold atmosphere component, while radiation coming from the radial (vertical) directions forms in the warm component. As the anisotropy is basically a measure of the excess of the radial radiation over the radiation from the side, this will lead to an increase in the anisotropy as compared to the warm model. The same effect decreases the anisotropy in colder areas near the boundary. However the anisotropy alteration scale in some frequency $\nu$ (which is a characteristic distance from the boundary to the regions where the considered effect can be neglected) is inversely proportional to the opacity at this frequency  (as these scales  correspond to the optical thickness one in both components). But the CN number density and, consequently, opacity strongly decrease in warm regions, so the increase in the anisotropy in such regions is significant over a greater geometrical scale than the decrease in cold regions. {  The considered effect will increase the mean anisotropy and therefore polarization as the anisotropy behavior at all depths has the same tendency: the total mean anisotropy is larger than the mean of the individual atmospheres. The illustration of the anisotropy behavior close to the boundary between different atmospheric components is given in the available online Sect.~\ref{subsec:app}} The efficiency of this effect in atomic lines and the continuum is much weaker, since their opacity dependence on temperature is not so steep as for molecules.
 Let us notice that the same effect changes also the mean intensity and, consequently, the intensity part of the source function and the emitted Stokes $I$. However,  the relative impact on the mean intensity is much smaller than  relative changes of the anisotropy (as for the mean intensity  the sum of the intensities coming from the radial and side directions is important, while for the anisotropy it is their difference). Moreover, the intensity source function contains also the thermal part (see Eq.~(\ref{eq:totalS})) while the polarization arises only in the scattering process. 

If the photon mean free path is much smaller than the characteristic size of  alternating cold and warm areas, the considered boundary effect is not important, since the anisotropy in most of the area is equal to the asymptotic value. If on the contrary, the photon mean free path is big, the horizontal radiation comes from many alternating parts of the atmosphere and the effect is smoothed out. Therefore, the considered anisotropy variations are most efficient when the photon mean free path is comparable with the granulation scale.

{  The mean photon free path in the layers with the strongest contribution to Stokes $Q$ is about 300--500 km and, therefore, is of the same order of magnitude as the granulation scale.} So, one can expect that the anisotropy alteration could be quite significant in CN lines. However,  in spite of the appropriate photon mean free path, this effect is not important for the continuum, as it only weakly depends on temperature. Therefore, an increase in the anisotropy in the warm atmosphere is almost exactly compensated by its decrease in the cold component.

Let us notice that the above consideration is valid only for optically thick lines. The CN lines under consideration remain optically thick even in the warmer atmosphere component. In the optically thin case the scattering polarization becomes proportional to the molecular number density and, therefore, the contribution from  warm regions reduces. 
This situation was discussed by \citet{trujilloetal2004} who found that the scattering polarization in optically thin ${\rm C}_2$ lines mainly comes from  colder upflowing regions. 

The considered effect can be the reason of our failure to reproduce the observations with a two-component atmosphere model in Sect.~\ref{subsec:multi}. 
Such modeling underestimates the anisotropy and, therefore, scattering polarization. It is important to notice that this conclusion is independent on the real granulation model. It is not even essential for us which part of the solar atmosphere is warmer at the line formation height: granular upflow or intergranular downflow. The origin of the reversed granulation in the solar atmosphere and the place where it occurs is still not fully clear \citep[see, for example][and references therein]{cheungetal2007}.

It appears that a self-consistent 3D modeling can improve the situation with the anisotropy. But even current advanced 3D models may not be able to provide the correct anisotropy as they involve many free parameters and assumptions. Thus, their ability to reproduce the CN observations analyzed in this paper could be a test for their reliability.

\subsection{Anisotropy correction in 1D modeling}\label{subsec:correction}
To account for the underestimated anisotropy we  artificially increase it by introducing a weighting function. Namely, in calculations of the second moment we multiply the intensity with a function $\eta(\mu)$ only within the Stokes $Q$ component of the source function,
\begin{eqnarray}
J_0^2(x)  = \! \! \!\!&& \frac{1}{4 \sqrt{2}} \int\limits_{-1}^{+1} 
\left (  (3 \mu'^2-1) I (x, \mu') \cdot  \eta(\mu')  \right.   \nonumber \\
 && \left. +  3 (\mu'^2-1) Q(x', \mu') \right )   d \mu',
\label{eq:qx_mod}
\end{eqnarray}
which is chosen to overweight the radiation coming from the radial (vertical) directions and give a lower weight to the radiation coming from aside. We used the following expression
\begin{equation}
\eta(\mu)=\frac{3(1+f \mu^2)}{3+f},
\end{equation}
where $f$ is a free parameter. Higher values of $f$ imply stronger corrections, while $f \! \! \!  =  \! \!0$ corresponds to the uncorrected anisotropy. Such a choice 
of function is quite convenient as in the case of zero anisotropy our correction obeys photon conservation. In general the error caused by additional photons remains on the order of the anisotropy, i.e. a few percents and is thus negligible because we did not introduce any correction to the zero-order moment of intensity (see Eq.(\ref{eq:jx})). Note that when applying the correction factor $\eta(\mu')$ also to Stokes $I$ component it would be essential to normalize the scattering integral. We did not introduce any correction to the zero-order moment of intensity (see Eq.~(\ref{eq:jx})) and also kept the intensity fixed during the POLY iterations.
With the purpose to avoid any additional differential effects we kept the parameter $f$ fixed in the atmosphere down to 150 km and then linearly decreased it to zero at $-100$ km. Examples of the anisotropy behavior with depth for several $f$ values are given in Fig.~\ref{fig:anis} for the case of the FALC atmosphere model. One can see that already $f \!\! = \! \!0.3$ value can solve the problem of negative anisotropy at the bandhead frequency, although as will be discussed in Sect.~\ref{sec:newmodeling} for better fit we need a somewhat higher value.

\section{Spectral CN modeling with anisotropy correction}\label{sec:newmodeling}
In the previous section we showed that introducing the new parameter $f$ into  
Eq.~(\ref{eq:qx_mod}) can solve the problem of the radiation field anisotropy deficiency. In this section we are employing again the standard FALC atmosphere model but with the $f$ parameter to regulate the anisotropy. Apart from this anisotropy parameter, there are two other main free parameters: the fraction of thermal photons at the temperature minimum layer $\delta_{\rm th}^{\rm min}$ and the magnetic field strength $B$. 
Variation of the first parameter affects all lines in more or less the same way, although some differential effects can be induced by  different line  formation depths and by the dependence of collisional rates  on height. Effective  Land\'{e} factors of the upper states for the CN violet system lines are inversely proportional to the angular momentum $N$, and therefore the magnetic sensitivity is decreasing with $N$ (see Eqs.~(\ref{eq:WH}--\ref{eq:gamma})). Thus, the Hanle effect acts on CN lines {\it differentially} which greatly reduces the model dependence of the deduced field. For more details see the discussion in \citet{shapiroetal2007b}.  

In our model the turbulent single-value magnetic field is assumed to be constant in the whole atmosphere. In Sect.~\ref{subsec:11} we will discuss the possibility to confirm or exclude some variability with the height.

\subsection{Region of the (0-0) bandhead}\label{subsec:00}
Here we present our results for the spectral region close to the (0,0) bandhead  (3880.5--3883.6 {\AA}). {  We found that for values $f \ge 0.4$ the observation can be fitted with a good quality.}

\begin{figure*}
\centering
\includegraphics{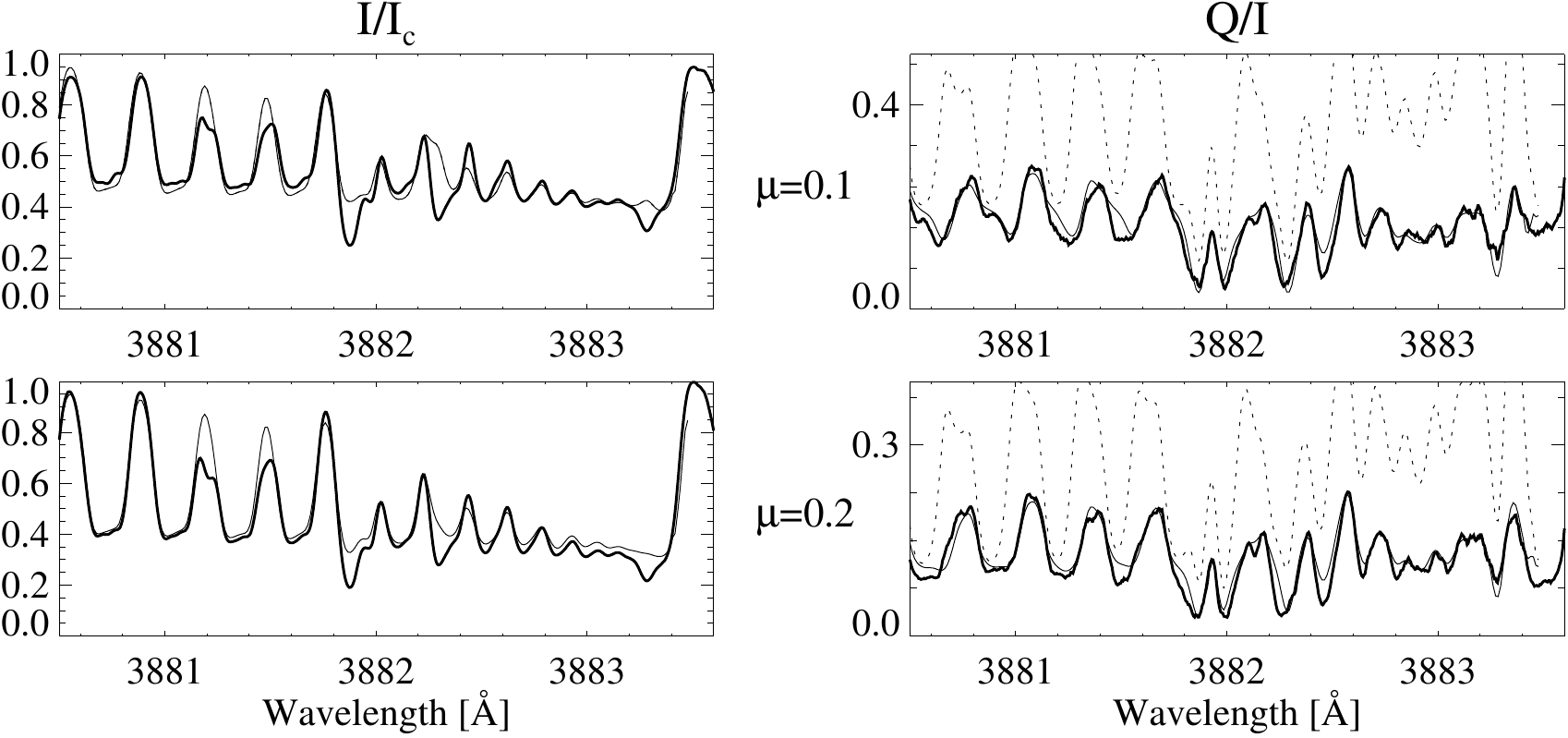}
\caption{Observations of Stokes $I/I_{\rm c}$ and $Q/I$  in the region of the (0,0) bandhead and fits obtained with the FALC atmosphere model and increased anisotropy. The three curves presented in each panel correspond to observations (thick solid) and to the calculations with zero magnetic field (dotted) and with a {  magnetic field strength of $82$ G (thin solid)}. Note that no differences between the calculated curves are apparent in intensity.}
\label{fig:IQ3884}
\end{figure*}

\begin{figure}
\resizebox{\hsize}{!}{\includegraphics{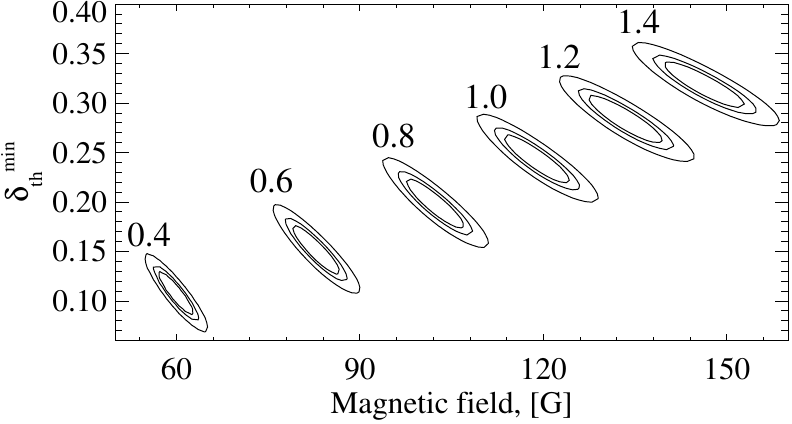}}
\caption{$\chi^2$-contours for solutions in the region of the (0,0) bandhead. The ellipses correspond to discrete values of $f$ indicated at the top. Three contours are drawn for each $f$ corresponding to confidence levels  68.3\%, 90\%, and 99.73\%.}
\label{fig:contour00}
\end{figure}

In Fig.~\ref{fig:IQ3884} we show one of the good solutions  yielding a magnetic field strength of {  $B=82$ G, a collisional coefficient $\delta_{\rm th}^{\rm min}=0.3$ and an anisotropy parameter $f=0.6$}. One can see that both $Q/I$ and $I/I_{\rm c}$ signals can be fitted well except for a few wavelengths affected by atomic blends. We recall here that background atomic lines were calculated only in LTE. Comparing the observed and zero-field $Q/I$ profiles one can see that the Hanle effect significantly depolarizes the spectrum at both $\mu$ values. Moreover its action leads to an apparent wavelength shift of four polarization peaks between 3880.6 {\AA} and 3881.8 {\AA}. The reason for this is that  these peaks are formed by pairs of doublets with angular momentum $N$ about 50 and 10, respectively, and in each peak the doublet with smaller $N$ (higher Hanle sensitivity) has a shorter wavelength and forms a blue part of the peak. Therefore, the blue part of each peak is  more strongly affected by the Hanle effect than the red one leading to the apparent shift. Such differential line behavior in the magnetic field allows us to disentangle the polarization changes caused by the collisional coefficient $\delta_{\rm th}^{\rm min}$ and magnetic field $B$  and, hence, to determine these parameters. 

However, increasing the coefficient $f$, which is in charge of the additional anisotropy, will lead to an overall growth of polarization. Such a  growth can also be achieved by decreasing the collisional coefficient. Thus, it is difficult to distinguish these effects with observations at only two different limb angles as adjustments of both parameters  influence polarization in approximately the same way (see, however, the discussion in Sect.\ref{subsec:11}, where we are dealing with observations at five different $\mu$).  The $I/I_{\rm c}$ spectra are not so sensitive to the $\delta_{\rm th}^{\rm min}$ coefficient and, therefore, cannot be used as an additional constraint. Finally, the effect of the increased  coefficient $f$ can be compensated by  increasing  the collisional coefficient $\delta_{\rm th}^{\rm min}$  and by increasing the magnetic field strength $B$. Therefore, we have executed each $\chi^2$-minimization with a fixed $f$ parameter.

To estimate the quality of the fit we use the $\chi^2$ concept and define
\begin{equation}\label{eq:chi}
\chi^2 = 
  \frac{1}{n}
\sum_{i=1}^{n}{ \left( \frac{ (Q/I)_\mathrm{obs} 
                              - (Q/I)_\mathrm{th} }
                            { \sigma_{Q/I} }  \right)^{2} },  
\end{equation}
where $n$ is the number of data points. For constraining our free parameters we used only Stokes $Q/I$   as intensity is not very sensitive to variations of $f$ and $\delta_{\rm th}^{\rm min}$  and  the line depths depend on the line blanketing parameters which can be slightly adjusted (see Sect.~\ref{sec:obs}). To compute $\chi^2$ we used 120 data points chosen under the condition to avoid blending with atomic lines. 

The $\chi^2$-contours for different $f$ parameter values are presented in Fig.~\ref{fig:contour00}.
For each $f$ value we have drawn three contours which correspond to the confidence limits  68.3\%, 90\%, and 99.73\%. Of course, the errors in our case are not necessarily distributed normally and there are definitely some systematic errors.
The deduced magnetic field strengths corresponding to  $\chi^2$ minima are listed in Table~\ref{table:bestfit}. {  A minimum $f$ parameter of $0.4$ is required for simultaneous fits to $I/I_{\rm c}$ and $Q/I$.} {  The minimum $\chi^2$ value corresponds to the magnetic field strength $B=82 \pm 10$ G.}

The main conclusion from  Fig.~\ref{fig:contour00} and Table~\ref{table:bestfit} is that the magnetic field strength appears to be model dependent. A larger additional anisotropy leads to a stronger field because the collisional rates enter into the denominator of  Eq.~(\ref{eq:gamma}).
It implies that their increase will lead to  weaker Hanle sensitivity, so for obtaining the same Hanle depolarization a higher field is needed. Moreover, the overall polarization growth can also be partially compensated by the Hanle depolarization (however, only within a limit which is constrained by the differential behavior of the Hanle effect). For a fixed $f$ value both the magnetic field and collision rate are well constrained. Even though decreasing the collisional rate (and hence increasing scattering) can be partly compensated by the growth of Hanle depolarization, the magnetic field strength is constrained due to the differential behavior of the Hanle effect. Knowing collisional rates more precisely would help to obtain a unique solution.
  
\subsection{Region of the (1-1) bandhead}\label{subsec:11}
\begin{figure*}
\centering
\includegraphics{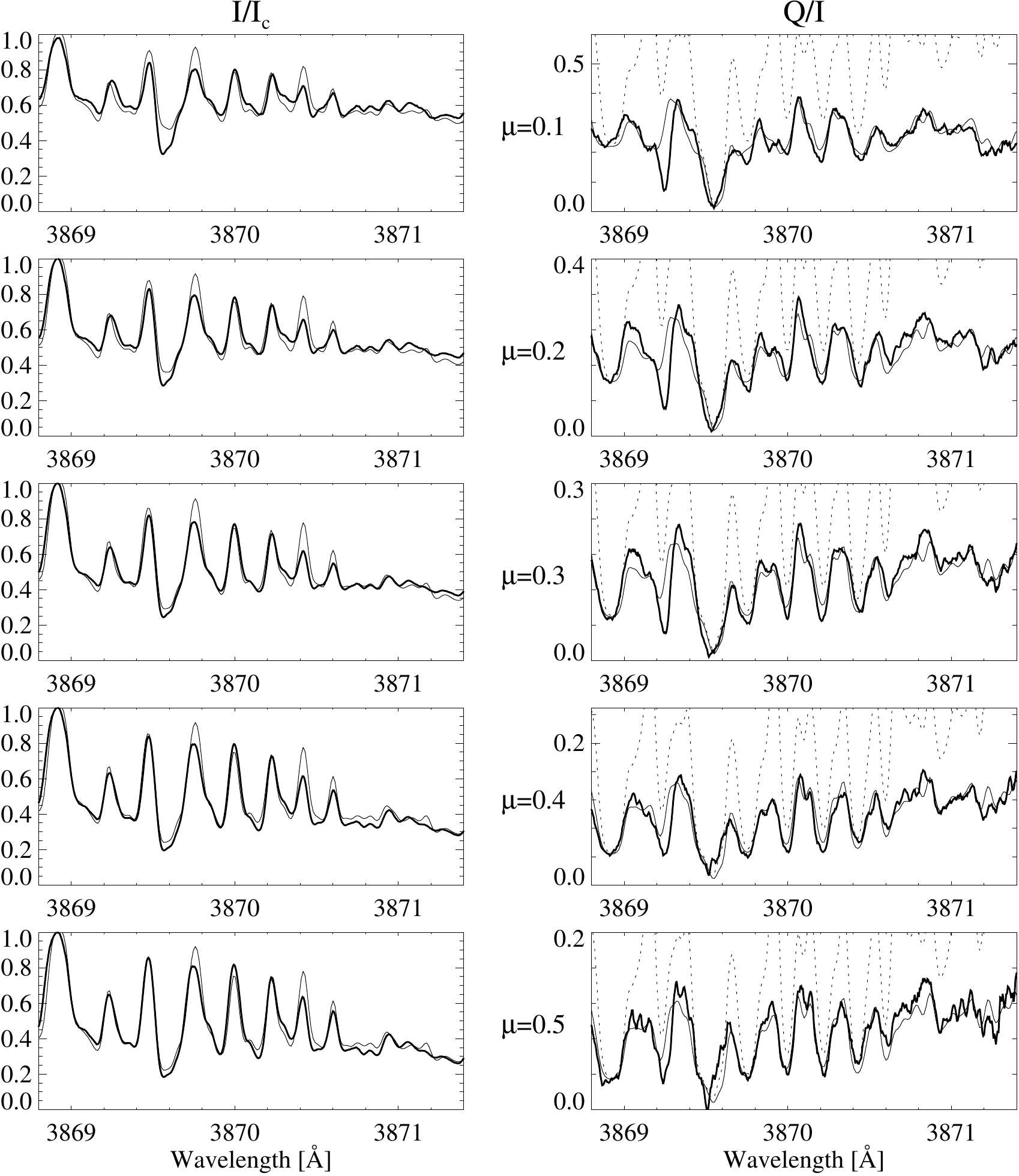}
\caption{Observations (thick solid lines) of Stokes $I/I_{\rm c}$ and $Q/I$  in the region of the (1,1) bandhead and fits obtained with FALC atmosphere model and increased anisotropy. Two curves presented in each $Q/I$ picture  correspond to the calculations with zero magnetic field (dotted) and with {  magnetic field strength of $45$ G (thin solid).}}
\label{fig:IQ3872_1}
\end{figure*}
\begin{figure}
\resizebox{\hsize}{!}{\includegraphics{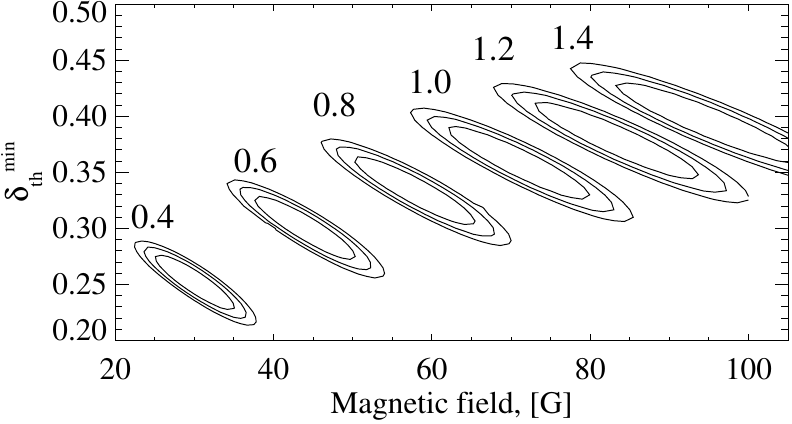}}
\caption{The same as Fig.~\ref{fig:contour00} for the (1,1) bandhead.}
\label{fig:contour11}
\end{figure}
Here we present our results for the spectral region close to the (1,1)  bandhead (3868.8--3871.5\,{\AA}). A comparison of the observed and calculated  spectra with {  a magnetic field strength $B=45$ G, collisional coefficient $\delta_{\rm th}^{\rm min}=0.16$ and an anisotropy parameter $f=0.6$  are shown in Fig.~\ref{fig:IQ3872_1}}. This region contains many CN lines from both (1,1) and (0,0) bands and also several strong atomic blends (among them two strong iron lines  at about 3869.5\,{\AA}) as well as one CH line at about 3870.1\,{\AA}. Therefore, the overall quality of the fit is worse than that for the (0,0) bandhead region. The $\chi^2$ minimum values and corresponding magnetic field strengths are listed in  Table~\ref{table:bestfit}, the $\chi^2$-contours are shown in Fig.~\ref{fig:contour11}. {  The minimum $\chi^2$ value corresponds to the magnetic field strength  $B=45 \pm 15$ G.}

The observations with higher $\mu$ values sample deeper layers of the solar atmosphere where the collisional rates and, consequently, the denominator of Eq.~(\ref{eq:gamma}) are larger. Therefore, these lines are less affected by the Hanle effect and less depolarized.
This can be clearly seen in Fig.~\ref{fig:IQ3872_1}: for $\mu=0.1$ the $Q/I$ curve is strongly depolarized in presence of a {  magnetic field strength of 45 G}, while for $\mu=0.5$ it is almost unaffected by the Hanle effect. The height dependence of the collisional rates and the deduced magnetic field  are strongly coupled with each other. For example a steeper collision rate dependence on temperature (and hence height) will lead to magnetic field strength. decreasing with height. Thus taken the current uncertainties in collisional rates and radiation field anisotropy it is not possible to unambiguously determine the depth dependence of the magnetic field strength. Improvements in collision  theory, 3D modeling of the radiation field anisotropy, and observations at a greater number of limb positions would be required for better constraints of the model atmosphere and the free parameters.

\begin{figure}
\resizebox{\hsize}{!}{\includegraphics{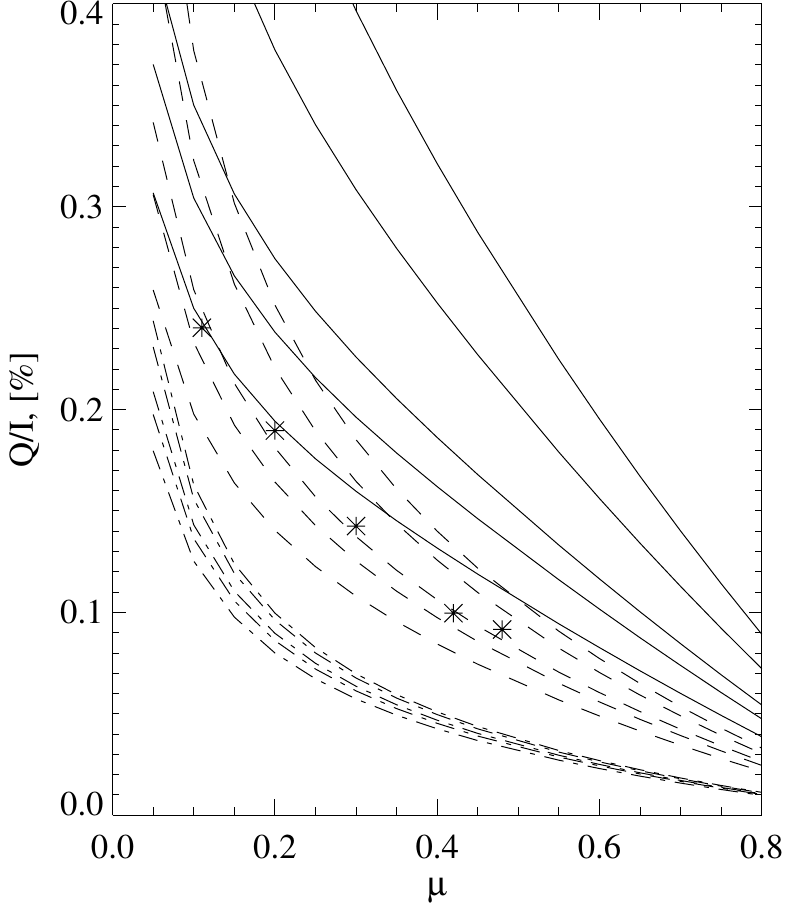}}
\caption{The dependence of the mean polarization {  in the (1,1) bandhead region} on the $\mu$ value for several collisional coefficients $\delta_{\rm th}^{\rm min}=$ 0.05 (solid lines), 0.3 (dashed lines), 0.7 (dashed dotted lines), and magnetic field strengths: 0\,G, 20\,G, 45\,G, 60\,G, and 90\,G (from top to bottom) for each given value of $\delta_{\rm th}^{\rm min}$. The five stars correspond to the five observed mean values.}
\label{fig:dep}
\end{figure}

Interestingly, there is one more mechanism which constrains the deduced magnetic field strength for this region. The alteration of the $\delta_{\rm th}^{\rm min}$ coefficient affects $Q/I$ curves at  all $\mu$ values in approximately the same way. The magnetic field  however mainly depolarizes lines observed at small $\mu$. This differential behavior significantly helps to constrain both the magnetic field strength and the collisional coefficient (assuming that we employ the correct dependency of the collisional rates on depth). {  In the (1,1) bandhead region} the considered mechanism is even more important than the differential Hanle effect as this region contains a crowded mixture of lines with different $J$ numbers, and the differential line behavior in the magnetic field is not clearly seen. This mechanism is illustrated in Fig.~\ref{fig:dep} where the dependence of the mean polarization on $\mu$ is plotted for several values of the collisional coefficient $\delta_{\rm th}^{\rm min}$ and magnetic field strength $B$.
The observed five points best coincide with the curve which corresponds to $B=45$ G and $\delta_{\rm th}^{\rm min}=0.3$.

From Figs.~\ref{fig:contour00} and \ref{fig:contour11} and Table~\ref{table:bestfit} one can see that the magnetic field strength  deduced with the same $f$ parameter for both spectral regions differ by a factor of two, with the one for the (0,0) bandhead being larger. This can be a signature of spatial magnetic field variations as the observations of these spectral regions were made on different days and sample accordingly different regions on the solar surface. We also interpreted the data from the atlas by \citet{gandorfer2005atlas} for the (0,0) bandhead region and found a weaker magnetic field of about 40 G compared to the 82 G obtained from our new observations with $f=0.6$. However, since the applied constraints are sensitive to the amount of available data (e.g., limb angles) only simultaneous measurements at the same limb distances in both regions can clarify the situation.

\begin{table}
\caption{$\chi^2$-minimum values and the least squares fits to the magnetic field strength $B$ for several $f$ parameter values.} 
\label{table:bestfit}
\centering 
\bigskip
\begin{tabular}{ c | c  c   |   c  c } 
\hline
\hline
\multicolumn{5}{c}{~~~~~~~~~~(0,0) band region~~~~~~~~~~~(1,1) band region }          \\
  $f$     & $\chi_{\rm min}^2$ &   $H$, G &    $\chi_{\rm min}^2$ &   $H$, G  \\ 
\hline 
     0.2 &  3.12   &      -          &  5.03 &  -             \\
   0.4 &  1.12   &      $59 \pm 10 $          &  2.26  &       $30 \pm 15$          \\
   0.6 &  0.98   &  $82 \pm 10 $   &  2.06 &     $45 \pm 15$ \\
   0.8 &  1.07   &  $102 \pm 15 $   &  2.25  &   $57 \pm 15$  \\
   1.0 &  1.24   &  $119 \pm 30 $   &  2.50  &   $70 \pm 15$  \\
   1.2 &  1.42   &  $133 \pm 30 $  &  2.72  &   $82 \pm 15$  \\
   1.4 &  1.61   &  $146 \pm 30 $  &  2.91  &   $105 \pm 15$  \\
  \hline 
\end{tabular}
\end{table}

\section{Conclusions}\label{sec:conc}
We have employed a new computational scheme to interpret simultaneously center-to-limb variations of the intensity and linear polarization  in the CN violet system. An analysis of these  variations improves our understanding of the scattering polarization in the CN violet system.  In our model we solve the statistical equilibrium equations and self-consistently account for multiple scattering in  optically thick molecular lines.  The use of the  Hanle effect allows us to probe unresolved entangled magnetic fields. 

We found that the standard FALC model is able to correctly reproduce intensity center-to-limb variations. However it fails to fit the linear polarization, in particular close to the CN bandhead where the calculated  polarization  becomes even negative (perpendicular to the solar limb). Moreover, we have shown that simultaneous fits to the intensity and linear polarization center-to-limb variations are not possible in 1D modeling, neither for one- nor for multi-component atmosphere models.   
Models which can provide the correct intensity center-to-limb variations are not able to produce enough anisotropy of the radiation field  to fit the linear polarization. 

We discuss several physical mechanisms which can increase the anisotropy. One of them is due to  
horizontal	temperature fluctuations in the atmosphere. This is especially valuable for molecular lines due to the strong dependency of the molecular number density  on temperature.

We account for the underestimated anisotropy  in 1D modeling by introducing an empirical weighting function into the radiative transfer. With this method we are able to fit center-to-limb variations of the intensity and polarization  simultaneously. The magnetic field can be determined via the differential Hanle effect but its value depends on the applied anisotropy correction.

The uncertainty in the fitted parameters such as the magnetic field is strongly connected to simplifications and shortcomings in the currently available theory of molecular collisions. {  Increasing the anisotropy in 1D models requires higher depolarization, which  will raise the low limits of collisions and magnetic fields. The same effect can occur in 3D models, if the anisotropy is overestimated, or vice versa.

Scattering polarization provides additional and independent observational constraints for testing the validity  of model atmospheres, not only in the case of 1D models but in particular also for testing results of 3D MHD simulations.

\begin{acknowledgements}
We are grateful to Han Uitenbroek for providing the original version of the RH-code. This work was supported by SNF grants 20002-103696, 200020-117821, and PE002-104552. SB acknowledges the EURYI Award from the ESF.
IRSOL is financed by the canton of Ticino, the city of Locarno together with CISL, and ETH Zurich.
\end{acknowledgements}


\bibliographystyle{aa}


\Online
\begin{appendix}
\section{Anisotropy variations}\label{subsec:app}
An example of anisotropy variations at the CN bandhead frequency due to horizontal temperature variations is given in Fig.~\ref{fig:boundary} for a very simple, purely illustrative model. We considered two adjacent homogeneous and isothermal atmospheres. One with temperature 4300 K (marked ``cold'') and another with temperature 4900 K (marked ``warm''). Such a choice gives a realistic temperature difference at line formation heights \citep{uitenbroek2000}.  The densities in both atmospheres were chosen under the condition that they are in the hydrostatical equilibrium with the FALC model at the height of 500 km (where the bandhead radiation at $\mu=0.1$ is formed). The anisotropy behavior  is drawn for four horizontal planes: at the surface (depth $D=0$ km), at the distance of one photon mean free path from the surface in the cold atmosphere ($D=200$ km), at the distance of one  photon mean free path from the surface in the warm atmosphere ($D=2000$ km), and deep in the atmosphere ($D=10000$ km). We indicated in boldface the anisotropy behavior in the narrower region of 1000 km as this roughly corresponds to the granulation scale.
Far from the boundary the anisotropy asymptotically reaches the limits individually calculated for each atmosphere (shown with dashed lines).  
At the surface, the asymptotic values are zero for both atmospheres (as the temperature gradient is zero), but the actual anisotropy is positive in the warm atmosphere and negative in the cold one. The behavior of the anisotropy in the warm atmosphere
is opposite to that in the cold one, while the mean anisotropy is positive.  At the depth of 200 km both  asymptotic limits become negative because of the surface effect. At the depth of 2000 km the asymptotic value for the cold atmosphere increases again to zero, as the layer above is optically thick and the surface effect can be neglected, while for the warm atmosphere the asymptotic value is still negative. And finally for the depth of 10000 km both asymptotic values reach again zero. 
 
\begin{figure}
\resizebox{\hsize}{!}{\includegraphics{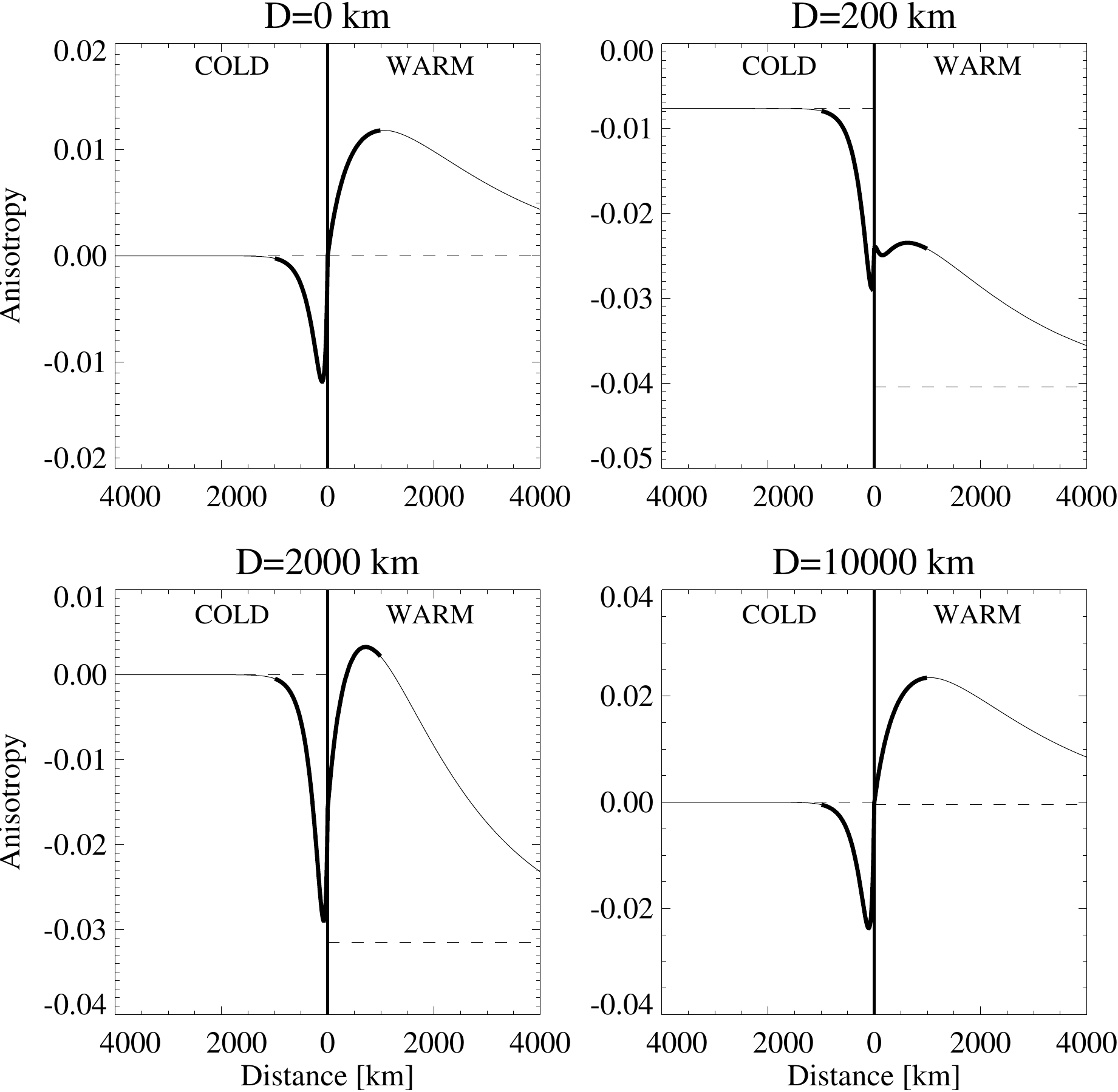}}
\caption{The dependence of the anisotropy on the distance from the boundary between cold and warm atmospheres. In each panel the cold atmosphere is shown to the left, the warm atmosphere is shown to the right. $D$ indicates the depth from the top of the semi-infinite homogeneous atmospheres.}
\label{fig:boundary}
\end{figure}

\end{appendix}


\end{document}